\documentclass{emulateapj-rtx4}
\usepackage{apjfonts}
\usepackage{graphicx}
\usepackage{color}

\def\mbh{{M_{\bullet}}}
\def\dotmbh{\dot{M}_{\bullet}}
\def\dotmacc{\dot{M}_{\rm acc}}

\def\rd{{d}}
\def\lax{{$\mathrel{\hbox{\rlap{\hbox{\lower4pt\hbox{$\sim$}}}\hbox{$<$}}}$}}
\def\gax{{$\mathrel{\hbox{\rlap{\hbox{\lower4pt\hbox{$\sim$}}}\hbox{$>$}}}$}}

\slugcomment{To appear in the Astrophysical Journal}
\shortauthors{LI, WANG \& Ho.}

\begin{document}

\title{Cosmological Evolution of Supermassive Black Holes. II. 
       Evidence for Downsizing of Spin Evolution}

\author{
Yan-Rong Li\altaffilmark{1}, Jian-Min Wang\altaffilmark{1,2} 
and Luis C. Ho\altaffilmark{3}
}

\altaffiltext{1}
{
Key Laboratory for Particle Astrophysics, Institute of High 
Energy Physics, Chinese Academy of Sciences, 19B Yuquan Road, 
Beijing 100049, China; liyanrong@mail.ihep.ac.cn, 
wangjm@mail.ihep.ac.cn 
}

\altaffiltext{2}
{
National Astronomical Observatories of China, Chinese 
Academy of Sciences, 20A Datun Road, Beijing 100020, China
}
\altaffiltext{3}
{
The Observatories of the Carnegie Institution of Washington, 
813 Santa Barbara St., Pasadena, CA 91101, USA; 
lho@obs.carnegiescience.edu
}

\begin{abstract} 
The spin is an important but poorly constrained parameter for describing 
supermassive black holes (SMBHs). Using the continuity equation of SMBH number 
density, we explicitly obtain the mass-dependent cosmological evolution of the 
radiative efficiency for accretion, which serves as a proxy for SMBH spin. Our 
calculations make use of the SMBH mass function of active and inactive 
galaxies (derived in the first paper of this series), the bolometric 
luminosity function of active galactic nuclei (AGNs), corrected for the 
contribution from Compton-thick sources, and the observed Eddington ratio 
distribution.  We find that the radiative efficiency generally increases with 
increasing black hole mass at high redshifts ($z\gtrsim1$), roughly as 
$\eta\propto M_\bullet^{0.5}$, while the trend reverses at lower redshifts, 
such that the highest efficiencies are attained by the lowest mass black 
holes.  Black holes with $M_\bullet\gtrsim10^{8.5}\,M_\odot$ maintain 
radiative efficiencies as high as $\eta\approx0.3-0.4$ at high redshifts, near 
the maximum for rapidly spinning systems, but their efficiencies drop 
dramatically (by an order of magnitude) by $z\approx0$.  The pattern for lower 
mass holes is somewhat more complicated but qualitatively similar.
Assuming that the standard accretion disk model applies, we suggest that the 
accretion history of SMBHs and their accompanying spins evolve in two distinct 
regimes: an early phase of prolonged accretion, plausibly driven by major 
mergers, during which the black hole spins up, then switching to a period of 
random, episodic accretion, governed by minor mergers and internal secular 
processes, during which the hole spins down.  The transition epoch depends on 
mass, mirroring other evidence for ``cosmic downsizing'' in the AGN 
population; it occurs at $z \approx 2$ for high-mass black holes, and 
somewhat later, at $z \approx 1$, for lower-mass systems.
\end{abstract}

\section{Introduction}
It has been realized since the pioneering work of \cite{Soltan82} 
that supermassive black holes (SMBHs) located at the centers of 
galaxies assemble their mass predominantly through accretion, an inference that 
has been further reinforced from considerations of the cosmic X-ray background 
(\citealt{Fabian99, Elvis02, Marconi04}). However, how SMBHs are fueled 
remains an outstanding unsolved issue. 

The spin of SMBHs traces the angular momentum of the accreted material. 
As such, it can serve as a powerful cosmic probe of SMBH feeding. 
However, spin is highly elusive to measure since its general 
relativistic effects emerge in the very vicinity of the horizon
(typically within a few tens gravitational radii). According to 
standard accretion disk theory, the radiative efficiency of energy 
conversion is closely linked to black hole spin 
through the marginally stable orbit, which is 
a function of the spin. The binding energy of the material in the accretion 
disk is locally radiated, and, after crossing the marginally stable orbit, 
the material freely falls into the hole without losing further energy 
due to the torque-free condition there. As a result, the total amount
of energy converted into radiation is the binding energy between
the marginally stable orbit and infinity (\citealt{Thorne74}).
Specifically, the radiative efficiency increases monotonically with black 
hole spin (e.g., see Figure 1 of \citealt{Martinez_Sansigre11a}).
This link allows us to analyze 
the net angular momentum of the accreted gas by quantifying 
the radiative efficiency, and hence obtain clues on how SMBHs are fed. 
By presuming a fixed radiative efficiency, \cite{Soltan82} 
connected the mass growth rate of SMBHs with the luminosity 
function (LF) of active galactic nuclei (AGNs). Subsequent 
studies explored the cosmic growth of SMBHs 
by comparing the local SMBH mass density with the integrated 
energy density of AGNs across time (e.g., \citealt{Chokshi92, Small92, 
Yu02, Marconi04, Shankar04, Cao08, Cao10}; see also a 
review of \citealt{Shankar09}). 
These studies found that an average radiative 
efficiency of $\eta\approx 0.1$ yields a local mass density
consistent with observational constraints.
From the theoretical 
point of view,  if the angular momentum of the accreted material 
stays aligned with the spin axis of of the black hole and the direction 
remains unchanged, the black hole will be rapidly spun up 
to a maximally rotating system and attain a radiative efficiency  of
$\eta\approx 0.42$ (\citealt{Thorne74}). The inconsistency between this 
estimate of $\eta$ and that based on So{\l}tan's argument potentially 
intimates that episodic transitions of angular momentum of accreted 
material during the active phases of black holes (i.e. random accretion) may 
play an important role in the cosmological evolution of SMBHs (e.g.,
\citealt{King08, Wang09}). It is thus expected that the spin of SMBHs, and 
hence their radiative efficiency, evolves with redshift. 

\cite{Wang09} constructed a formalism to determine the evolution of the
radiative efficiency, which depends completely on observables. When 
applying this formalism to survey data, they found that 
SMBHs are spinning down with cosmic time since $z\le 2$. Such 
an evolutionary trend of spins suggests that SMBH growth is 
driven by episodic random accretion. Since different SMBH 
populations may undergo different accretion histories, it is 
important to verify whether the spin evolution depends on black hole 
mass. This can give further insights into how SMBH activity is 
triggered and how the angular momentum of the accreted material 
ultimately influences SMBH spin.

Deep surveys in the X-rays and in other bands have established that AGN 
activity exhibits cosmic downsizing (e.g., \citealt{Ueda03, Bongiorno07,
Hopkins07, Cirasuolo10}). Generally, the space density of AGNs 
with low luminosity peaks at lower redshift than 
that of AGNs with high luminosity. Two distinct scenarios 
can result in such luminosity evolution. If black holes are 
accreting at near-Eddington luminosity once they become active, 
downsizing can be caused by activity shifting toward low-mass black 
holes at low redshift (e.g., \citealt{Heckman04, Shankar09, Schulze10}).
Alternatively, downsizing can result from a decrease of the average accretion 
rate onto black holes at low redshift (\citealt{Babic07}); in other words, 
the entire black hole population shines with high Eddington ratio at high 
redshift and then slowly fades out over time.  These two scenarios can be
tested using information on black hole demographics. 
Studies of local AGN samples unambiguously find that local 
active black holes are typically an order of magnitude less massive 
than the typical inactive black holes residing in normal galaxies 
(\citealt{Heckman04, Greene07}). Using the virial method to measure 
the black hole masses of SDSS quasars out to redshift 
$z\approx 4$, \cite{Labita09a, Labita09b} show that the maximum mass  
of the active black hole population notably increases with redshift. 
These observations seem to suggest that cosmic 
downsizing arises from activity shifting over mass. 
Recalling that the angular momentum carried into the black hole
along with mass accretion changes the black hole spin,
we may expect that the spin evolution of black holes 
would follow a behavior similar to that of the redshift evolution of AGN 
activity.

Assuming that SMBHs grow mainly through accretion, we
extend the study of Wang et al. (2009) to explicitly
show the mass-dependent behavior of the cosmic evolution of SMBH spin.
In Section 2, we construct a generalized 
equation that expresses the radiative efficiency as a function of black hole 
mass and redshift. Section 3 shows the Eddington ratio distribution used
in our calculations, and Section 4 derives the SMBH mass function of 
AGNs, including Compton-thick sources. We then present the results for 
SMBH growth in Section 5 and SMBH spin evolution in Section 6.
The implications of our results on accretion scenarios are discussed 
in Section 7.  Conclusions are summarized in Section 8.

Throughout the paper, we adopt a cosmological model with 
$\Omega_{\rm M}=0.3$, $\Omega_\Lambda=0.7$, and 
$H_0=70\,{\rm km~s^{-1}~Mpc^{-1}}$. 

%
\begin{figure*}[t]
 \centering
 \includegraphics[angle=-90.0, width=0.8\textwidth]{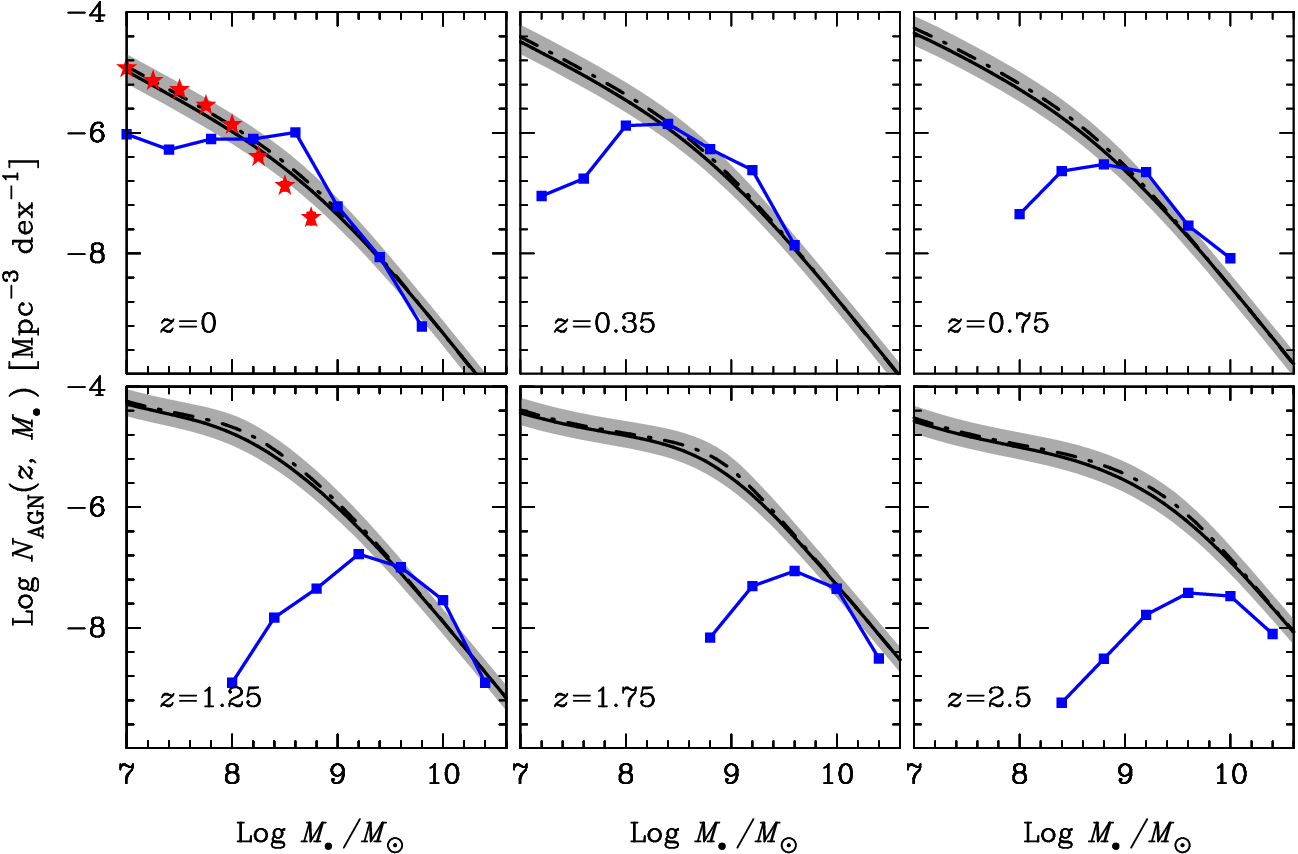}

 \caption{SMBH mass function of AGNs. Solid and dot-dashed lines 
 are the calculated mass function using Equation (\ref{equ_nagn}) 
 with a log-normal and power-law distribution of Eddington ratios, 
 respectively. The shaded areas represent a typical error of
 $\Delta\log N_{\rm AGN}\approx\pm0.2$ dex from the AGN bolometric LF of
 \cite{Hopkins07}. Asterisks in the $z=0$ 
 bin  are from \cite{Greene07}, 
 and squares are from \cite{Vestergaard09}, who also compiled 
 the SMBH mass function at $0.025<z<0.5$ from the Bright Quasar Survey.
 For the sake of a comparison, we superpose it in the $z=0$ bin. 
 Note that the turnover toward the low-mass end of the mass function 
 of \citeauthor{Vestergaard09} is due to high survey incompleteness 
 (\citealt{Kelly10}). }

\label{fig_agnmf}
\end{figure*}
%

%
%
\section{The Continuity Equation of Black Holes}
Let $N(t, \mbh)$ be the SMBH mass function, including both active and inactive 
black holes, which specifies the number of black holes per unit comoving 
volume and per unit mass at cosmic time $t$. Its evolution is described by a 
continuity equation (e.g., \citealt{Small92})
\begin{equation}
    \frac{\partial N(t, \mbh)}{\partial t}
  + \frac{\partial }{\partial\mbh}
    \left[N(t,\mbh)\langle\dotmbh\rangle\right]
  = S(t, \mbh),
\end{equation}
where $\langle\dotmbh\rangle$ is the mean mass accretion rate
for inactive and active SMBHs with mass $\mbh$ at time $t$, and 
$S(t, \mbh)$ is the source term that accounts for black hole mergers. 
As in most previous works (e.g., \citealt{Small92, Yu02, Marconi04}), we 
neglect black hole mergers, setting $S(t, \mbh)=0$.
\cite{Shankar_etal09, Shankar10} investigated the importance of black 
hole mergers on the evolution of the SMBH mass function and concluded 
that the effect of mergers is minor compared with mass accretion. 
Numerical simulations by \cite{Volonteri05} and \cite{Berti08} also 
verified that mass accretion dominates over mergers in determining 
the mass growth and spin distribution of black holes.

The radiation power of a SMBH with mass accretion rate $\dotmacc$ 
is $L=\eta\dotmacc c^2$. Because of radiative losses, the mass 
growth rate of the SMBH is $\dot M_{\bullet}=(1-\eta)\dotmacc$, 
yielding
$\dot M_{\bullet}=(1-\eta)\lambda L_{\rm Edd}/\eta c^2$.
Here the Eddington ratio is defined as $\lambda=L/L_{\rm Edd}$, 
where $L_{\rm Edd}=4\pi G\mbh m_pc/\sigma_{\rm T}$, $G$ is the 
gravitational constant, $m_p$ is the proton mass, $c$ is the light 
speed, and $\sigma_{\rm T}$ is the Thomson electron scattering 
cross section. With the help of the duty cycle of SMBHs, $\delta(t, \mbh)$, 
the mean mass accretion rate can be written as
\begin{equation}
   \langle\dot M_{\bullet}\rangle
  =\delta(t,\mbh)\dot M_\bullet
  =\delta(t,\mbh)\frac{1-\eta}{\eta}
   \frac{{\bar \lambda}(t,\mbh) L_{\rm Edd}}{c^2},
\end{equation}
where $\bar\lambda(t, \mbh)$ is the mean Eddington ratio determined 
from observations (see Equation \ref{equ_mean_ratio} below). 
The duty cycle is usually defined as 
\begin{equation}
  \delta(t, \mbh)=\frac{N_{\rm AGN}(t, \mbh)}
                  {N_{\rm G}(t, \mbh) + N_{\rm AGN}(t, \mbh)},
\end{equation}
where $N_{\rm G}(t, \mbh)$ and $N_{\rm AGN}(t, \mbh)$ are the SMBH mass 
functions of galaxies and AGNs, respectively, and, accordingly 
\begin{equation}
N(t, \mbh)=N_{\rm G}(t, \mbh)+N_{\rm AGN}(t, \mbh).
\end{equation}
Combining the above equations, we rewrite the continuity equation
\begin{equation}
 \frac{\partial N(z,\mbh)}{\partial z}=-\frac{\rd t}{\rd z}
      \frac{\partial}{\partial \mbh} \left[\frac{1-\eta}{\eta}
      \frac{\bar\lambda(z,\mbh)L_{\rm Edd}N_{\rm AGN}(z, \mbh)}{c^2}
      \right],
\label{equ_con}
\end{equation}
where and hereinafter we substitute the cosmic time $t$ with the 
corresponding redshift $z$. Integrating Equation (\ref{equ_con}) 
over $M_{\bullet}'$ from $\mbh$ to $\infty$, we obtain
the radiative efficiency
\begin{equation}
\eta^{-1}(z, \mbh)=1+\frac{c^2}{\dot{u}(z,\mbh)}
    \left(\frac{\rd t}{\rd z}\right)^{-1} 
    \frac{\partial }{\partial z}
    \int^\infty_{\mbh}N(z,M_\bullet')\rd{M_\bullet'},
\label{equ_eta}
\end{equation}
where the AGN luminosity density is
\begin{equation}
\dot{u}(z,\mbh)=\bar\lambda(z,\mbh) L_{\rm Edd} N_{\rm AGN}(z,\mbh),
\label{equ_udot}
\end{equation}
and we use the boundary condition 
$\dot{u}(z,\mbh\rightarrow\infty)=0$.  
Equation (\ref{equ_eta}) is the generalized $\eta-$equation of 
Wang et al. (2009). The underlying rationale of this generalized 
$\eta-$equation lies in the assumption that mass accretion only increases the 
SMBH mass, but keeps the SMBH number density conserved. Subsequently, 
variations of the SMBH number density in the interval $[\mbh, \infty]$ 
should come from the ``inflow'' of SMBHs through the boundary 
$M_\bullet'=\mbh$ due to mass accretion. As a result, we obtain the 
radiative efficiency for SMBHs with any given mass of $\mbh$ from 
the AGN luminosity density.

To solve Equation (\ref{equ_eta}), it is imperative to obtain three 
ingredients that appear on the right-hand side: the SMBH mass functions 
of normal galaxies and AGNs, and the Eddington ratio distribution. 
In the first paper of this series (\citealt{Li11}, Paper I),  
we have derived the SMBH mass function of normal galaxies out to $z\approx2$ 
using the latest galaxy luminosity and stellar mass functions.  In what 
follows, we show in detail how to obtain the SMBH mass function of AGNs from 
the observed Eddington ratio distribution.

%
\begin{figure}[t!]
\centering
\includegraphics[angle=-90.0, width=0.42\textwidth]{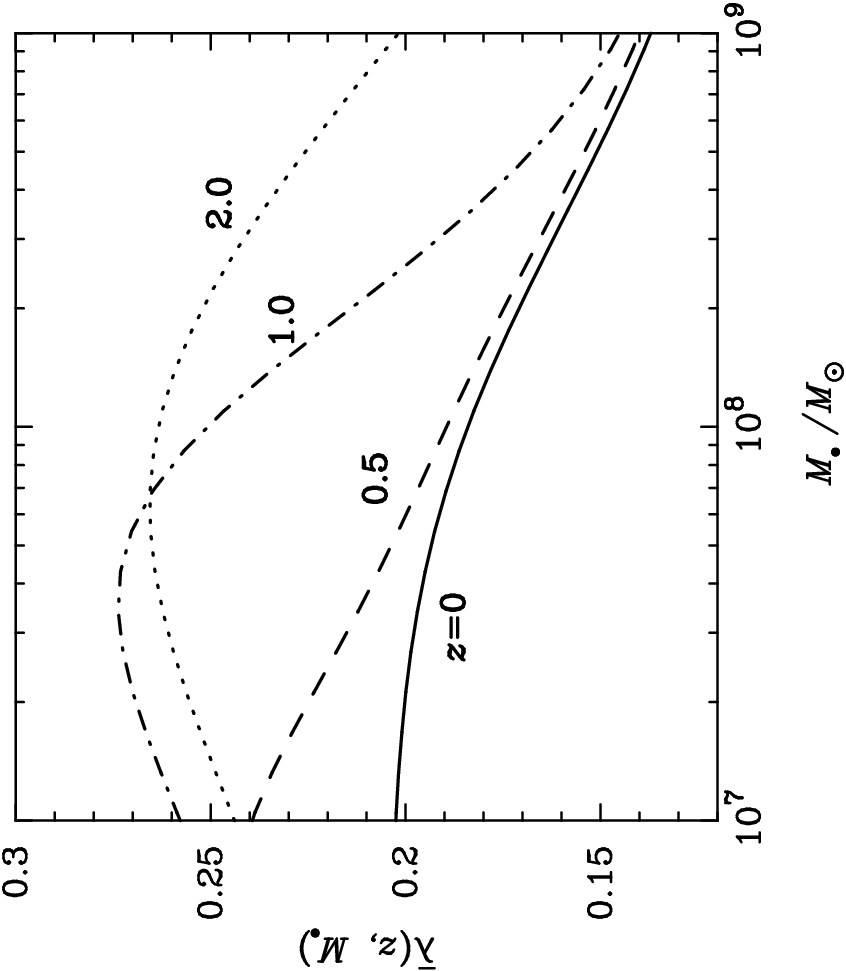}\\
\includegraphics[angle=-90.0, width=0.40\textwidth]{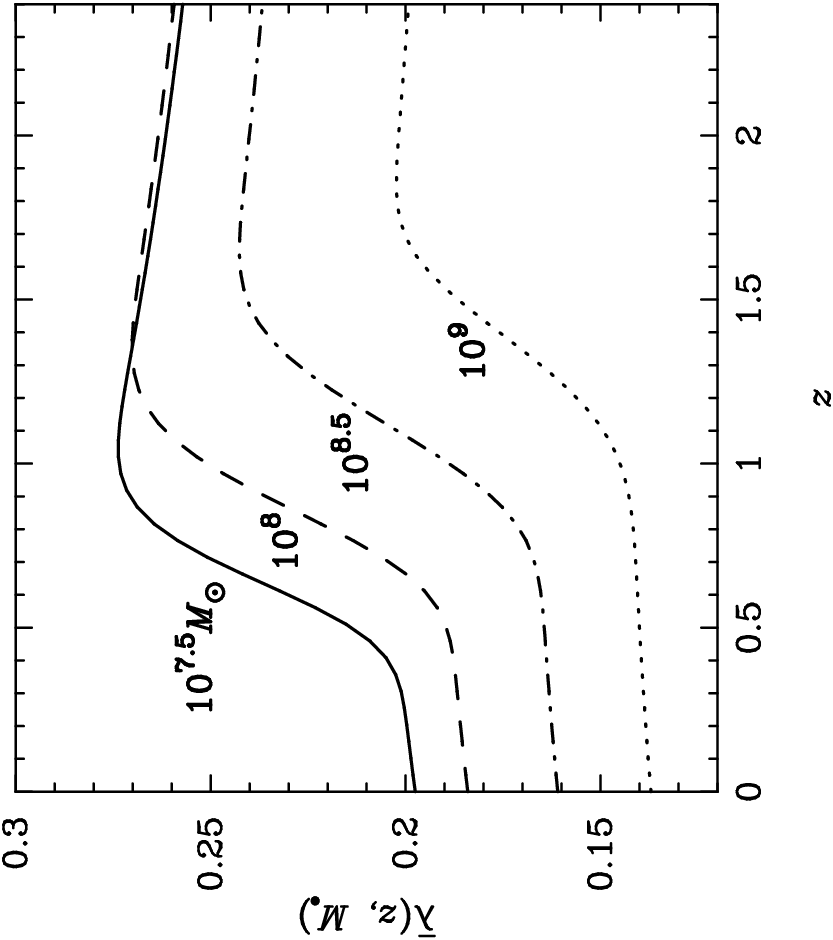}

\caption{Mean Eddington ratio as a function of (top) black hole mass and 
(bottom) redshift.}
\label{fig_lambda}
\end{figure}
%

%
%
\begin{figure}[t!]
\centering
\includegraphics[angle=-90.0, width=0.42\textwidth]{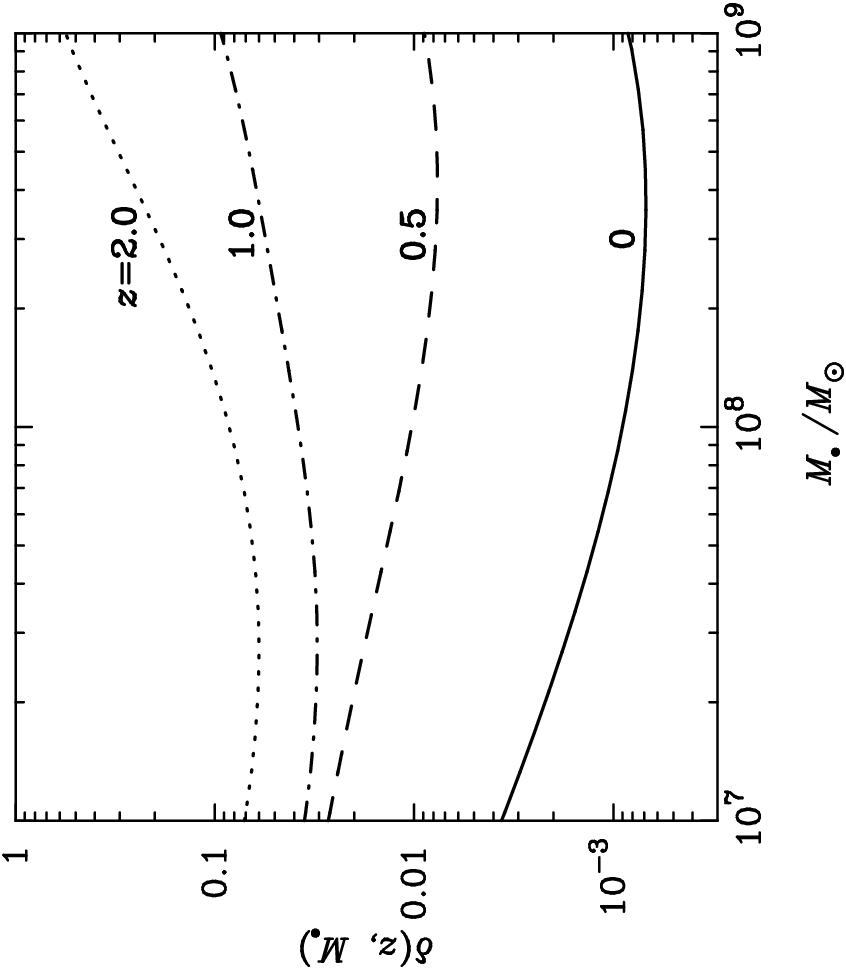}
\includegraphics[angle=-90.0, width=0.40\textwidth]{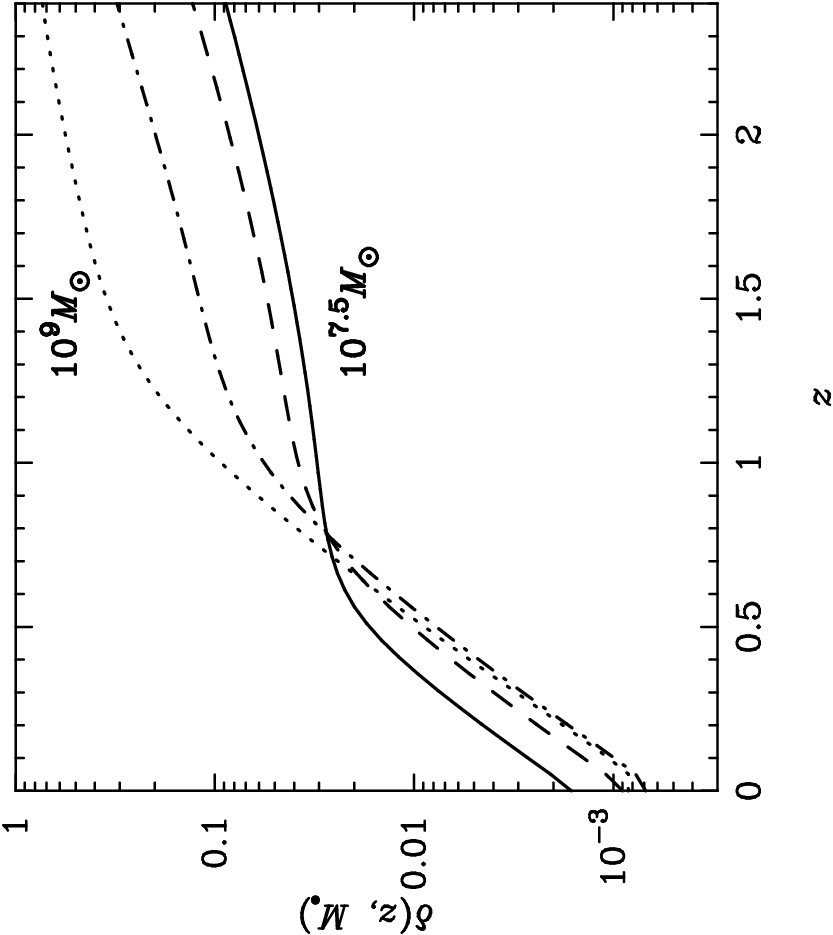}

\caption{Duty cycle as a function of (top) black hole mass and (bottom) 
redshift.}
\label{fig_duty_cycle}
\end{figure}

\section{Eddington ratio distributions}
Observations thus far have not yet reached a consensus on the 
observed Eddington ratio distribution, especially at high 
redshifts. The challenges arise from systematic uncertainties 
in measuring black hole mass using the virial relation and the 
flux limit of surveys.  However, among earlier studies, two 
types of Eddington ratio distributions are preferred. Generally, 
the observed Eddington ratios derived from the bright AGN (quasar) 
samples exhibit a log-normal distribution (\citealt{Kollmeier06, 
Shen08, Kauffmann09, Kelly10}). By contrast, those derived from 
samples that include fainter sources or low-luminosity AGNs display
a power-law distribution and/or an additional log-normal component in the 
high-Eddington ratio regime (\citealt{Heckman04, Yu05, Hopkins09, 
Kauffmann09, Schulze10}). In particular, the peak and dispersion
of the log-normal distribution are found to be almost independent of 
luminosity and redshift (\citealt{Kollmeier06, Shen08}). Through 
analysis of the properties of the host galaxies selected from 
SDSS, \cite{Kauffmann09} argued that in the log-normal regime 
black holes self-regulate their growth in environments with 
a plentiful supply of cold gas, so that the Eddington ratio is 
retained at a universal rate independent of luminosity 
(black hole mass) and the properties of the host galaxies. 
By contrast, in the power-law regime the gas has 
run out and the central black holes are probably being fueled by 
mass loss from stars in the bulges, thereby appearing as low-luminosity 
AGNs \citep{Ho09a, Ho09b}.

With the above lines of observations,  we employ two types of Eddington 
ratio distribution in our calculations. We express the log-normal 
distribution as
\begin{equation}
P(\lambda)d\log\lambda=\frac{1}{\sqrt{2\pi}\sigma}
     \exp\left[-\frac{(\log\lambda-\mu)^2}{2\sigma^2}\right]
     d\log\lambda,
\label{equ_lognormal}
\end{equation}
where $\mu$ is the logarithm of the peak Eddington ratio and $\sigma$ 
is the dispersion. Below we will show how to determine $\mu$. The dispersion 
$\sigma$ is found to be inessential to our results, and we set $\sigma=0.3$ 
(\citealt{Kollmeier06}). The power-law distribution is given by 
\begin{equation}
P(\lambda)d\log\lambda=C_0\left(\frac{\lambda}{\lambda_0}\right)^{-\kappa}
             \exp\left(-\frac{\lambda}{\lambda_0}\right)d\log\lambda
      ~~~~~~~\lambda>\lambda_{\rm min},
\label{equ_powerlaw}
\end{equation}
where $C_0$ is the normalization, $\lambda_0$ is the characteristic 
ratio, $\kappa$ is the power-law index, and $\lambda_{\rm min}$ is the
lower cut-off (\citealt{Hopkins09}; 
see also \citealt{Cao10}). According to their feedback-regulated 
model, \cite{Hopkins09} suggest typical values of 
$\lambda_0\approx 0.2-0.4$ and $\kappa\approx0.3-0.8$. In our 
calculations, we set $\lambda_0=0.3$ and $\kappa=0.6$ as fiducial 
values. Note that for the power-law distribution, a lower cut-off 
$\lambda_{\rm min}$ is required to completely determine the 
normalization $C_0$. We adjust the values of $\lambda_{\rm min}$ 
so that the calculated SMBH mass function of AGNs matches the 
observed one.

\cite{Cao08} derived the mean Eddington ratio for a given black hole 
mass and redshift from the observed Eddington ratio distribution. 
For clarity, we present the details here. The SMBH mass function 
of AGNs is calculated by combining the observed Eddington 
ratio distribution and the AGN bolometric LF as
\begin{equation}
N_{\rm AGN}(z, \mbh)=\int\Phi(z, L_{\rm bol})
                     \frac{\rd\log L_{\rm bol}}{\rd\log \mbh}
                     P(\lambda)\rd \log\lambda,
\label{equ_nagn}
\end{equation}
where $\Phi(z, L_{\rm bol})$ is the AGN bolometric LF. 
Applying Bayes' theorem, the Eddington ratio distribution 
for a given black hole mass $\mbh$ is
\begin{equation}
\omega(z,\lambda|\mbh)=\frac{\Phi(z,L_{\rm bol})P(\lambda)}
                            {N_{\rm AGN}(z,\mbh)}.
\end{equation}
The mean Eddington ratio of AGNs with $\mbh$ at redshift $z$ is
\begin{equation}
\bar\lambda(z,\mbh)=\int\lambda\omega(z,\lambda|\mbh)\rd\log\lambda,
\label{equ_mean_ratio}
\end{equation}
where $\int\omega(z,\lambda|\mbh)\rd\log\lambda=1$. 

In Figure \ref{fig_agnmf}, we compare with observations the SMBH mass function 
of AGNs calculated by Equation (\ref{equ_nagn}) using two types 
of Eddington distributions. To  delineate the evolution of the AGN populations,
we adopt Hopkins et al.'s 
(2007) bolometric LF, which combines a large set of AGN LF measurements
in the rest-frame optical, soft and hard X-ray, and near-IR and mid-IR bands,
spanning a range of bolometric luminosities from
$\sim10^{42}$ to $10^{49}~{\rm erg~s^{-1}}$.
\cite{Greene07} measured 
the local SMBH mass function of broad-line AGNs from SDSS 
using standard virial relations. Based on the same method,
\cite{Vestergaard09} compiled the SMBH mass function of broad-line
AGN samples drawn from the Large Bright Quasar Survey, 
the Bright Quasar Survey, and SDSS, covering a range of redshift 
from the local epoch up to $z=5$. Note that the turnover toward
the low-mass end of the mass function of \citeauthor{Vestergaard09}
is due to the incompleteness of the surveys. 
\cite{Kelly10} carefully estimated the incompleteness of the SDSS
sample of \citeauthor{Vestergaard09} and found that it is 
highly incomplete for $\mbh\lesssim10^9\,M_\odot$ at $z>1$. 
Since these measurements just 
cover the unobscured population of AGNs, a correction factor has to be applied
to Equation (\ref{equ_nagn}) to account for such a selection bias
before performing this comparison (see Section 4 below).  
By fine tuning the value of $\mu$ for the log-normal distribution 
of Eddington ratio, we find that $\mu=-0.6$ produces an AGN SMBH mass
function that is in excellent agreement with the observed ones, 
particularly for the data of \cite{Greene07} at $z=0$, as
shown in Figure \ref{fig_agnmf}. 
Similarly, a value of $\lambda_{\rm min}=0.1$ for the power-law
distribution also gives an almost identical mass function.
However, it is probably unphysical that the Eddington ratio for the
power-law distribution is cut off at $\lambda_{\rm min}=0.1$. 
Nevertheless, whichever type of Eddington ratio 
distribution is used has very little influence on the final results
in the sense that the resultant AGN SMBH mass functions are very insensitive 
to the actual choice (see Figure \ref{fig_agnmf}).
Hereafter, we only use the log-normal distribution in our calculations.

Figure \ref{fig_lambda} shows the dependence of the mean Eddington ratio on
black hole mass and redshift. We find that the mean Eddington 
ratio roughly spans values in the range $\sim0.1-0.3$, slightly increasing
with redshift and decreasing with black hole mass. These trends
between the mean Eddington ratio and black hole mass and redshift 
are due to the power-law shape of the AGN LF and the redshift 
evolution of the slope of the power law. At the same time, we know
that nearby massive active galaxies generally contain black holes 
accreting at an Eddington ratio substantially below unity 
(e.g., \citealt{Ho08, Ho09a}). This does not conflict with the present 
results, considering that local luminous AGNs are quite 
rare. Moreover, we here focus on 
efficient accretion phases with high Eddington ratios,  during which 
black hole growth predominately happens (e.g., \citealt{Hopkins06,Xu10}).

%
\begin{figure}[t!]
\centering
\includegraphics[angle=-90.0, width=0.42\textwidth]{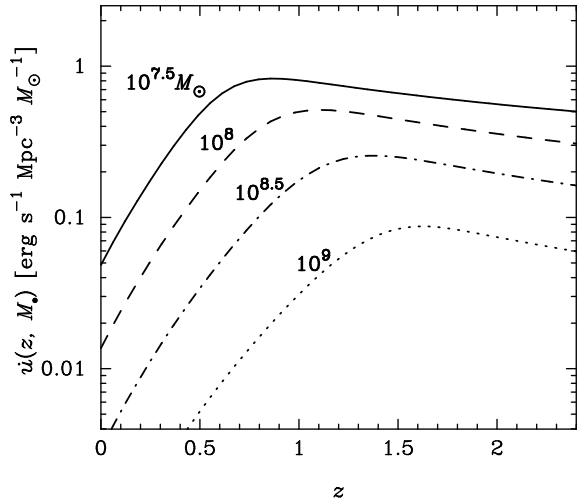}
\caption{Luminosity density $\dot u$ (see Equation \ref{equ_udot}) as a 
function of redshift, for different black hole masses.}
\label{fig_udot}
\end{figure}

%

%
\section{Accounting for Compton-thick AGNs}
Given the Eddington ratio distribution and AGN bolometric LF, 
it is trivial to calculate the SMBH mass function for AGNs using 
Equation (\ref{equ_nagn}). However, it is important to note that different 
bands are subject to different selection biases. Optical LFs miss 
obscured AGNs (e.g., \citealt{Bongiorno07}), while hard X-ray 
surveys ($\sim2-10$ keV) can trace only AGNs that are not completely
Compton-thick (e.g., \citealt{Hasinger08}), namely sources with column 
densities $N_{\rm H}\lesssim 10^{24}$ cm$^{-2}$.  \cite{Treister10} 
recently presented an evolution model for AGNs hosted by galaxies
undergoing major mergers and find that Compton-thick 
sources play an important role in the mass growth of SMBHs.  
Meanwhile, evidence is mounting from deep X-ray surveys, in combination
with IR surveys, that the population of Compton-thick AGNs with intermediate 
luminosities is of the same order as that 
of Compton-thin AGNs
(sources with $N_{\rm H}\approx 10^{22}-10^{24}{\rm~cm^{-2}}$;
e.g., \citealt{Daddi07, Alexander08, Fiore09, 
Treister09}). Population synthesis models of the cosmic X-ray background 
also predict the existence of a large number of Compton-thick AGNs 
(e.g., \citealt{Gilli07, Draper09}). 

How the fraction of Compton-thick AGNs evolves with redshift is not 
well known.  Even the overall evolution 
of Compton-thin AGNs is in debate. After correcting for
selection bias in their hard X-ray-selected AGN sample spanning 
the redshift range $z=0-4$, \cite{Treister06} find that the fraction of 
obscured AGNs increases with redshift as $(1+z)^{\alpha}$, with 
$\alpha\approx0.3-0.5$. \cite{Ballantyne06}, modeling the X-ray background,
also report a similar evolution trend of $(1+z)^{0.3}$.
However, other studies cast doubt on the redshift 
evolution of the obscured AGN fraction (\citealt{Ueda03, Dwelly06,
Gilli07, Lamastra08}). Based on their AGN population model, 
\cite{Gilli10} argue that by including a proper $K-$correction
an intrinsically constant obscured AGN fraction can also represent the data.

Motivated by the above results, we parameterize the redshift dependence of 
the fraction of Compton-thick for all AGNs as
\begin{equation}
f_{\rm C}=f_{\rm C, 0}(1+z)^{\alpha},
\label{equ_ct}
\end{equation}
with $f_{\rm C, 0}\approx0.3$ and $\alpha\approx0.3$. Here we 
determine $f_{\rm C, 0}$ by assuming that the local fraction of 
obscured (Compton-thin and Compton-thick)
to unobscured AGNs is $\sim3:1$ (\citealt{Treister06}), and that 
there is an equal abundance of Compton-thin  and Compton-thick sources.
As described later, our final results are quite insensitive to the value
of $\alpha$; therefore, we set $\alpha=0.3$, consistent with \cite{Treister06}
and \cite{Ballantyne06}.  On the other hand,  
it is well established that the obscured fraction of Compton-thin 
AGNs decreases with luminosity (\citealt{Ueda03,La_Franca05,Hasinger08}).
The dependence of obscuration on luminosity for 
Compton-thick AGNs, however, is still poorly understood. We neglect 
this effect, but we discuss the implications in Section 6.2.3.

%
\begin{figure}
\includegraphics[angle=-90.0, width=0.48\textwidth]{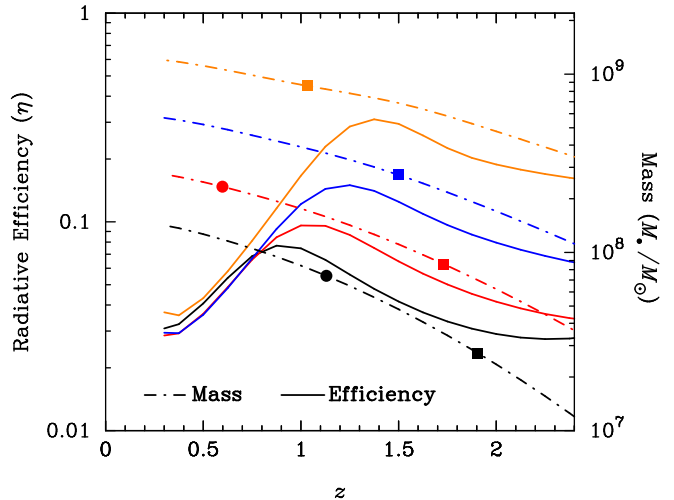}
\caption{Evolution of black hole mass (dot-dashed lines) and
 associated radiative efficiency (solid lines) as a 
 function of redshift. The initial masses are set to
 $10^{7}$, $10^{7.5}$, $10^{8}$, and $10^{8.5}$ $M_\odot$ 
 at redshift $z_0=2.5$. The symbols, squares and circles, 
 indicate when the black hole mass has $e$-folded once and twice,
 respectively.}
\label{fig_growth}
\end{figure}
%

%
\begin{figure*}[t!]
\centering
\includegraphics[angle=-90.0, width=0.75\textwidth ]{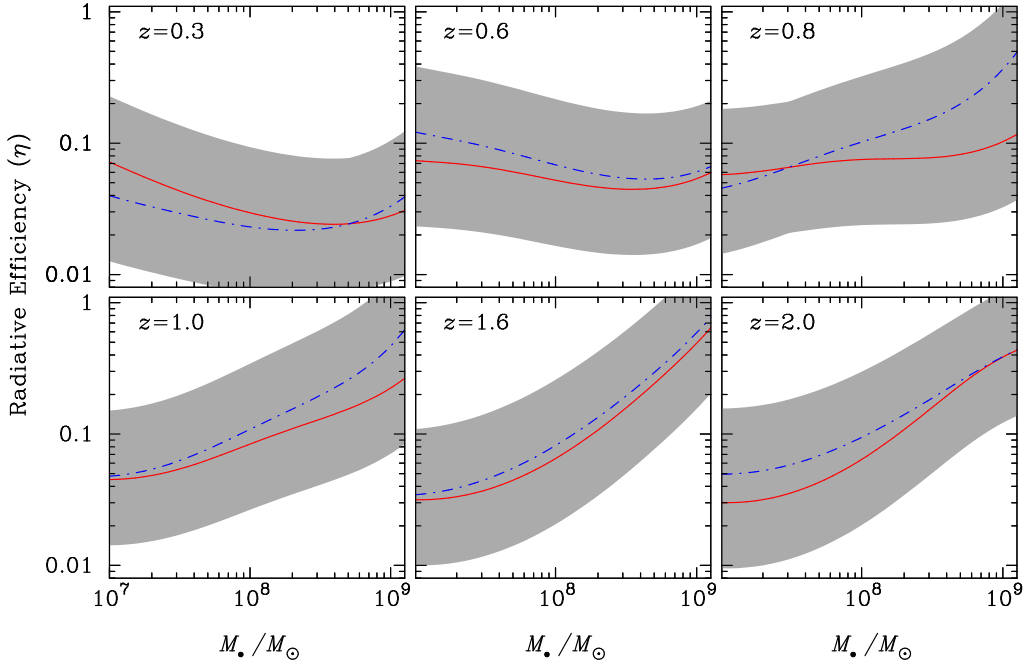}
\caption{Radiative efficiency as a function of black hole mass at different 
 redshifts. Solid lines and dot-dashed lines are the efficiency 
 using the SMBH mass function derived from the galaxy luminosity 
 and stellar mass function, respectively. Shaded 
 areas denote typical errors of $\Delta\log\eta=\pm$0.5 dex from 
 SMBH mass function and AGN LF. }
\label{fig_eta_mbh}
\end{figure*}
%

%
\begin{figure}[t!]
\centering
\includegraphics[angle=-90.0, width=0.45\textwidth ]{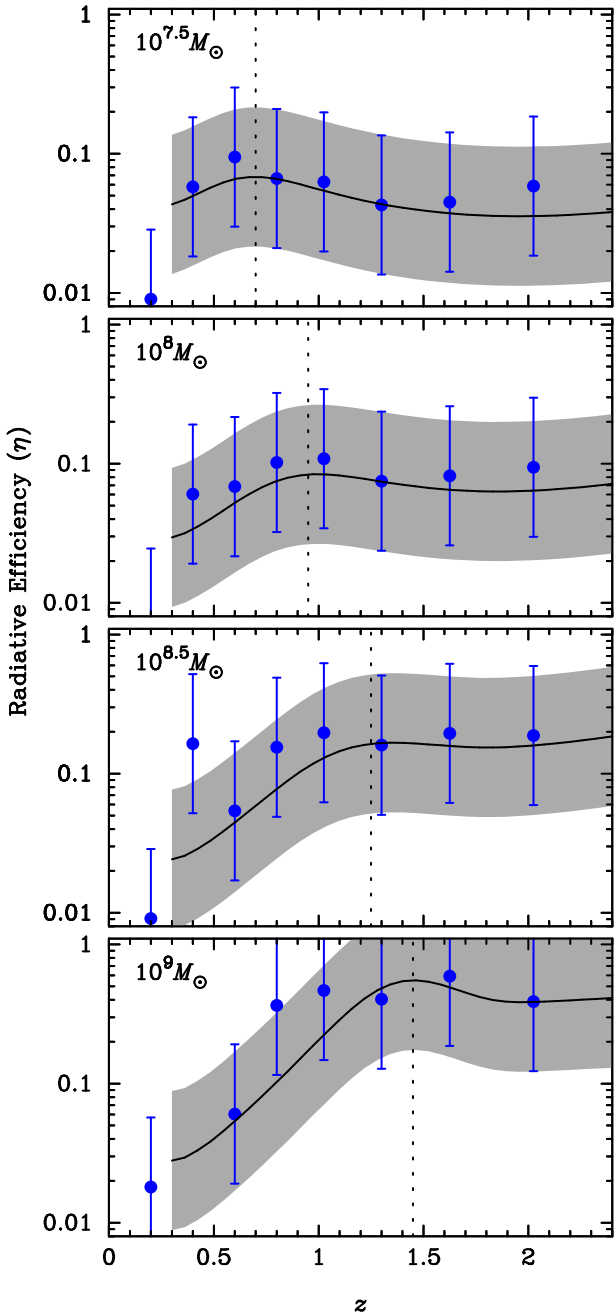}
\caption{Radiative efficiency evolution for different black hole masses. 
 Solid lines and data points are the efficiency using the SMBH mass 
 function derived from the galaxy luminosity and stellar 
 mass function, respectively. Shaded areas and error bars denote 
 typical errors of $\Delta\log\eta=\pm$0.5 dex from the SMBH mass 
 function and AGN LF. Vertical dotted lines represent the redshift 
 below which the efficiency begins to decline.}
\label{fig_eta_z}
\end{figure}
%

%
%
\section{Episodes of SMBH Growth}
With the AGN SMBH mass function determined from the previous sections, 
we can calculate the AGN duty cycle, as shown in Figure \ref{fig_duty_cycle}. 
At low redshifts, the duty cycle decreases with increasing black hole mass, 
indicating that the fraction of galaxies that are active also declines with 
mass (\citealt{Greene07, Shankar09}). Conversely, as a consequence of the 
cosmic downsizing of AGN activity, at high redshifts the duty cycle reverses 
trend and increases with black hole mass.
Duty cycles for $10^7\,M_\odot$ and 
$10^9\,M_\odot$ black hole are $\delta\approx4\times10^{-3}$ and  
$\sim8\times10^{-4}$ at $z=0$, respectively, but rise up to $\sim0.06$ 
and  $\sim0.5$ by $z=2$. As shown in the bottom panel of 
Figure \ref{fig_duty_cycle}, the active fraction of all black holes 
increases significantly from $\sim10^{-3}$ at $z=0$ to $\sim0.1$ 
at $z=2$, in concert with the feature of AGN LF that
displays a rapid rise up till $z=2$ (see also \citealt{Wang06}). 

With the duty cycle determined, the 
active time of AGNs is simply given by 
$\tau(z, \mbh)=\delta(z, \mbh) H(z)^{-1}$, where $H(z)$ is the 
Hubble parameter at redshift $z$ (\citealt{Peacock99}). 
As a result, the active time of AGNs is of the order of $\sim10^7$~yr 
in the local Universe and $\sim10^8$~yr at $z=2$.  This is consistent with 
the observational estimates from other methods (see \citealt{Martini04}
for a review). 

Figure \ref{fig_udot} illustrates the AGN luminosity density 
$\dot u(z,\mbh)$, defined by Equation (\ref{equ_udot}), and its variation
with redshift for different black hole masses. Since the mean Eddington 
ratio is approximately constant over black hole mass and redshift, 
$\dot u(z,\mbh)$ just reflects the evolution 
of the AGN SMBH mass function. Again, the behavior of cosmic downsizing 
is evident: more massive active black holes reach their 
maximum number density earlier than less massive ones.

With the mean accretion rate $\langle\dot M_\bullet\rangle$, 
we can trace the SMBH growth history since redshift $z_0$ as
\begin{equation}
\mbh(z)=M_0+\int_{z_0}^z \left\langle\dot M_\bullet \left[z',\mbh(z')
            \right]\right\rangle\frac{dt}{dz'}dz',
\end{equation}
where $M_0$ is the initial black hole mass at redshift $z_0$. 
Figure \ref{fig_growth} plots the growth history of black holes with initial 
masses $10^{7}$, $10^{7.5}$, $10^{8}$, and $10^{8.5}$ $M_\odot$
at $z_0=2.5$. The symbols (squares and circles) indicate when the black 
hole mass has $e$-folded once and twice.  For a hole with an initial mass of 
$10^7\,M_\odot$, it grows to $\sim1.5\times10^8\,M_\odot$ 
at $z\approx0.3$, $e$-folding roughly 3 times. 
A $10^{8.5}\,M_\odot$ hole grows just to
$\sim1.0\times10^9\,M_\odot$ at $z\approx0.3$, 
$e$-folding only once. 
\footnote{Our estimates of $e$-folding times and mass growth rates 
are qualitatively consistent with previous studies (e.g., \citealt{Marconi04, Merloni08}).  
Note, however, that these studies assumed a constant radiative efficiency.}
Recall that the $e$-folding time is quantified 
by the Salpeter (1964) time
\begin{equation}
t_{\rm Sal}=\frac{\eta}{(1-\eta)\bar\lambda}
            \frac{c\sigma_{\rm T}}{4\pi Gm_p}
           =2.3\times10^8\left(\frac{10\eta}{1-\eta}\right)
            \left(\frac{\bar\lambda}{0.2}\right)^{-1}{~\rm yr}.
\end{equation}
It is obvious that the $e$-folding time scales in proportion to the
radiative efficiency. 
The decline of radiative efficiency makes black hole growth possible,
especially for low-mass black holes at low redshift, considering 
that the AGN life time is of the order of $10^7$~yr at that epoch 
(see also the discussions of \citealt{King06, Netzer07, Netzer_etal07}).

An intriguing issue that has attracted much attention over the years
is what is the mass range of seed black holes and how they grow
to present-day SMBHs (e.g., \citealt{Volonteri08, Wang08}).  In principle, an 
extrapolation of the present approach to much higher redshifts, 
say $z\gtrsim10$, may be helpful to place constraints on the properties of seed
black holes. This is beyond the scope of the present paper, and we defer
this investigation to future work.

%
\section{SMBH Spins}

%
%
\subsection{Evolution of the Radiative Efficiency}

In Figure {\ref{fig_eta_mbh}} we present the radiative efficiency, calculated 
using Equation (\ref{equ_eta}), as a function of SMBH mass, using the SMBH 
mass function derived from both the galaxy LF and the galaxy stellar mass 
function. Shaded areas represent typical errors of $\Delta\log\eta=\pm0.5$ dex 
from the SMBH mass function ($\sim0.3$ dex) and AGN LF ($\sim0.2$ dex).
We find that at low redshifts (e.g., 
$z\approx 0.3$) SMBH of all masses are accreting material with a relatively low 
radiative efficiency. At high redshifts (e.g., $z\approx 1-2$), although the 
uncertainties are large, the radiative efficiency generally increases 
with black hole mass. A rough correlation 
is $\eta\propto M_\bullet^{\gamma}$, with $\gamma\approx 0.5$.
Indeed, \cite{Cao08} argued on the basis of the So{\l}tan argument 
that the radiative efficiency should increase with black hole 
mass so as to guarantee that the mass function of local relic SMBHs matches 
the mass function of galaxies. Previous studies, assuming a 
fixed radiative efficiency over black hole mass and redshift,
concluded that $\eta\approx0.1$ (e.g.,
\citealt{Yu02, Marconi04}). The values of the radiative efficiency 
we obtain here are consistent with this average value.

The redshift dependence of the radiative efficiency is plotted in 
Figure {\ref{fig_eta_z}} for different black hole masses. Solid lines 
show the results obtained using the SMBH mass function derived from the
galaxy LF, while the superposed data points make use of the SMBH mass function 
derived from the galaxy stellar mass function, computed for specific redshift 
bins. Again, shaded areas and error bars represent typical errors of $\Delta\log\eta=\pm0.5$ 
dex. For high-mass black holes ($M_\bullet\gtrsim10^{8.5}\,M_\odot$), the 
radiative efficiency maintains a high value of $\eta\approx 0.3$ at $z\approx 
2$, but then strongly declines toward lower redshifts.  The radiative 
efficiency for $\mbh\approx 10^9\,M_\odot$ at $z\approx 2$ marginally exceeds 
the maximum efficiency of $\eta\approx0.4$ allowed for extreme Kerr black 
holes.  This may arise from the unrealistic treatment of the mean Eddington 
ratio distribution beyond a black hole mass of $10^{9}\,M_\odot$. 
Another cause may be an overestimate of the correction factor 
for Compton-thick AGNs in Equation (\ref{equ_ct}), if the factor 
is generally anti-correlated with luminosity. 
For low-mass black holes, the evolution 
is more complicated. The radiative efficiency appears to plateau
at $z\approx 1-2$ and then decreases dramatically toward $z=0$.
Recently, \cite{Wang09} obtained the average 
radiative efficiency for black hole masses
$M_\bullet\gtrsim10^{8.2}\,M_\odot$ with redshift $z<2$. 
They found a similarly rapid decrease of $\eta$ 
from $z\approx 2$ to the local epoch (see their Figure 2a). 

Figure \ref{fig_growth} shows how the black 
hole mass and the associated radiative efficiency evolve
over time. Interestingly, there exists a peak 
in the efficiency curve: for a given initial mass at redshift $z=2.5$, 
the efficiency gently rises to a local maximum, and then sharply falls off.
The rise of the efficiency during black hole growth is due to the 
positive correlation between efficiency and mass, as shown in Figure 
\ref{fig_eta_mbh}, while the subsequent decline is due to the systematic 
decrease of the efficiency toward low redshifts. It seems that the 
efficiency evolution can be characterized by two regimes,  one during which 
the efficiency increases and another during which it decreases.

\subsection{Evolution of SMBH Spins}
If the standard accretion disk model applies to the AGNs under consideration, 
we can elucidate their spin evolution through the radiative efficiency 
obtained above (e.g., \citealt{Thorne74}). Since black holes gain their mass 
predominately during their quasar phase (e.g., \citealt{Hopkins06,Xu10}), the
standard accretion disk model is quite a reasonable approximation.  From 
Figure \ref{fig_eta_mbh}, it is apparent that at high redshifts ($z\gtrsim1$) 
black hole spin increases with mass. At low redshifts the dependence of spin 
on mass seems reversed, although large uncertainties prohibit us from reaching
a firm conclusion. 
Volonteri et al. (2007) have drawn a connection between black hole spin and 
galactic morphology and argued that SMBHs in elliptical galaxies possess
higher spins than those in spiral galaxies.  Such a morphology-related spin 
distribution has been used to explain the observed radio-loudness bimodality 
in nearby AGNs (e.g., \citealt{Sikora07}). Our results at $z\approx 0.3-0.6$ 
are not qualitatively consistent with this scenario.  One possible explanation 
for this inconsistency is that radio-loud galaxies comprise only a tiny 
fraction of the galaxy population, whereas our results pertain to a global 
average of all galaxies. Moreover, our current approach is unable to constrain 
the dispersion of the spin distribution at any given redshift or mass.

Most importantly, Figure \ref{fig_eta_z} shows that high-mass 
black holes begin to spin down earlier than low-mass black holes.
To guide the eye, the vertical dotted lines in the figure mark the redshift 
below which the efficiency begins to decline.  Recalling the cosmic downsizing 
of AGN activity, there appears to be a connection between black hole spin 
and accretion. Indeed, if random accretion plays a role 
in black hole growth, spin evolution inevitably follows 
cosmic downsizing just as AGN activity does, since high-mass black 
holes become active earlier and hence are spun down earlier by 
random accretion. However, we observe two regimes in the evolution of the 
radiative efficiency: SMBHs first spin up, and then down.  This
indicates that more complex accretion scenarios need to be explored.
We return to this point in Section 7.

%
%
\subsection{Influence of Uncertainties}
We explore the reliability of the radiative efficiencies in light of some of 
the assumptions used in our calculations.

%
\begin{figure}[t!]
\centering
\includegraphics[angle=-90.0, width=0.42\textwidth ]{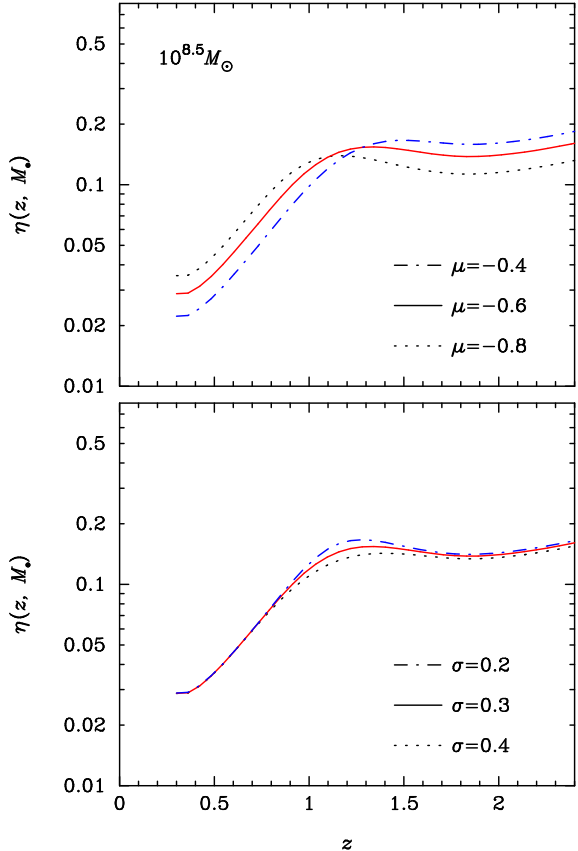}
\caption{Influences of the observed Eddington ratio distribution on radiative 
efficiency for black holes with mass $\mbh=10^{8.5}\,M_\odot$.  The 
dependence on the mean ($\mu$) and dispersion ($\sigma$) of the distribution 
is illustrated in the top and bottom panels, respectively.}
\label{fig_error_ratio}
\end{figure}
%

%
%
\begin{figure}[t!]
\centering
\includegraphics[angle=-90.0, width=0.42\textwidth ]{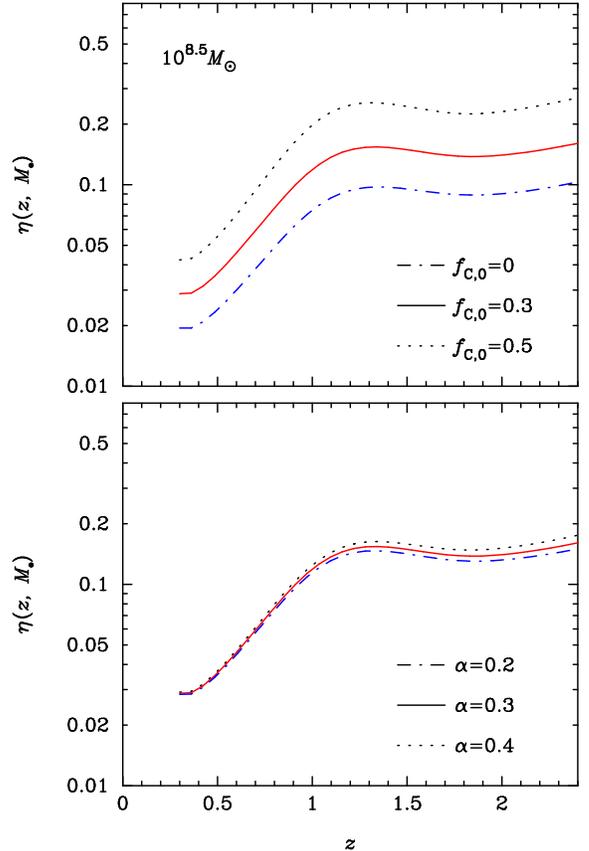}
\caption{Influences of the fraction of Compton-thick AGNs on 
   radiative efficiency for black holes with mass 
   $\mbh=10^{8.5}\,M_\odot$. The 
dependence on the normalization ($f_{\rm C,0}$) and redshift dependence 
($\alpha$) of the Compton-thick fraction (see Equation 13)
is illustrated in the top and bottom panels, respectively.}
\label{fig_error_CT}
\end{figure}

\subsubsection{SMBH Mass Function of Galaxies} 

The SMBH mass function is a crucial input because it determines the mass growth 
rate in Equation (\ref{equ_eta}). In Paper I, we derived the 
SMBH mass function of galaxies out to redshift $z\approx 2$ using 
the latest luminosity and stellar mass functions of field galaxies 
to constrain the masses of their spheroids. SMBH masses were inferred through 
a locally calibrated empirical correlation between black hole mass
and spheroid mass. The SMBH mass functions derived from the galaxy 
luminosity and stellar mass functions show very good agreement, 
both in shape and in normalization, testifying to the robustness 
of our results. After carefully examining various sources of 
uncertainties, we found that the total uncertainties on the 
SMBH mass function are generally within $\sim0.3$ dex, and possibly even 
smaller for lower mass black holes (see Paper I for details).
As a consequence, uncertainties in the SMBH mass function introduce an 
uncertainty less than $\sim0.3$ dex to the resultant radiative efficiency.

\subsubsection{Eddington Ratio Distributions} 

The observed distribution of Eddington ratios is described as 
log-normal, independent of luminosity and redshift 
(e.g., \citealt{Kollmeier06,Shen08}). Such a distribution may be
realistic in the context of a self-regulation model for accretion 
of black holes embedded in a gas-rich environment (\citealt{Kauffmann09}).
To assess its impact on our  results, we calculate the radiative
efficiency for three values of the mean Eddington ratio (logarithmic values
$\mu = -0.4, -0.6,$ and $-0.8$) and its dispersion ($\sigma$ = 0.2, 0.3, and 
0.4) for $\mbh = 10^{8.5}\,M_\odot$ (Figure \ref{fig_error_ratio}). We find 
that $\mu$ does have a mild impact on the redshift position where black hole 
spin begins to decline: smaller $\mu$ results in a later decrease of spin.
However, in comparison with Figure \ref{fig_eta_z}, the adopted 
uncertainty of $\Delta\mu=\pm0.2$ does not alter the overall 
cosmic downsizing behavior for the spin evolution.
Future systematic studies should try to 
directly constrain the intrinsic Eddington ratio distribution for a given 
black hole mass, instead of inferring it from the observed Eddington ratio 
distribution combined with the AGN LF. The dispersion of the log-normal 
distribution $\sigma$ has only a very minor influence on the radiative 
efficiency.

\subsubsection{Compton-thick AGNs}

We adopt a simple, redshift-dependent correction factor to account for
the selection bias of Compton-thick AGNs in Hopkins et al.'s (2007) bolometric 
LFs (Equation \ref{equ_ct}).  Its possible dependence on luminosity 
is neglected. From Equation (\ref{equ_eta}), it is apparent that a higher 
correction factor would enhance the energy density and therefore lead
to a larger radiative efficiency (see also \citealt{Martinez_Sansigre09}).
By analogy, if obscuration anti-correlates with luminosity, the radiative 
efficiency of low-mass black holes will be reduced.  This will influence the 
relation between spin and mass obtained in Figure \ref{fig_eta_mbh}. 
Unfortunately, the overall luminosity dependence of the fraction of 
Compton-thick AGNs remains unsettled still (e.g., see \citealt{Gilli10}), 
precluding us from applying a rigorous treatment.  Figure \ref{fig_error_CT} 
shows the influence of the correction for Compton-thick AGNs on the radiative 
efficiency for $10^{8.5}\,M_\odot$ holes. The results are significantly 
affected by the normalization factor $f_{\rm C,0}$, but they are quite 
insensitive to the redshift parameter $\alpha$. In any case, the overall 
evolution trend of radiative efficiency is qualitatively preserved.

\subsubsection{SMBH Mergers} 

As in most previous works, we neglect the term for black hole mergers in 
the continuity equation and, motivated by So{\l}tan's argument, assume that 
black hole growth is mainly driven by baryon accretion.  As mentioned above, 
previous studies show that mass accretion dominates over mergers in 
determining the mass growth and spin distribution of black holes 
(\citealt{Berti08,King08}). \cite{Shankar_etal09, Shankar10}, using 
theoretically predicted  merger rates from hierarchical structure formation 
models, also conclude that the effect of mergers is minor compared with mass 
accretion.  However, based on a semi-analytic model of hierarchical galaxy
formation and evolution incorporating black hole growth, \cite{Fanidakis11b} 
recently showed that mergers do contribute significantly to the final black 
hole mass for high-mass black holes (e.g., $\mbh\gtrsim10^9\, M_{\odot}$).  
Both of these calculations are subject to the uncertainties of the adopted 
recipes for merger rates of black holes, which are not well understood yet.

\subsection{Comparison with Previous Results}
Several recent parallel studies have attempted to observationally estimate 
SMBH radiative efficiency or spin. 
Directly starting from the definition of radiative efficiency, 
\cite{Davis11} obtained the efficiencies of individual sources in a 
sample of 80 Palomar-Green quasars using the bolometric luminosity 
and the accretion rate determined from accretion disk model spectral fits. 
They found an average $\eta\approx 0.1$ and a trend of increasing 
efficiency with increasing black hole mass (see also 
\citealt{Raimundo11}). The redshift range of their sample is 
confined to $z\leq0.5$.  The correlation of the radiative efficiency with 
black  hole mass found by \cite{Davis11} is not completely compatible with our 
results (i.e.  redshift bins $z=0.3$ and $0.6$ in Figure \ref{fig_eta_mbh}),
although both calculations suffer from large uncertainties.

Another common but highly model-dependent way to constrain 
SMBH spin is through the jet/radio power of radio-loud AGNs. In a series of 
works, \cite{Daly09a, Daly09b, Daly11} analyzed the 
radio beam power using large samples of extended radio sources 
and then determined the black hole spin with the aid of a
correlation between the beam power and spin.
Their results generally suggest that black hole spin in
radio sources decreases moderately from redshift $z\approx2$ 
to the local Universe.  Conversely, assuming that 
the efficiency of jet production is uniquely dependent on 
black hole spin, \cite{Martinez_Sansigre11a} inferred the spin
distribution by modeling the radio luminosity function
of radio sources with high- and low-excitation narrow 
emission lines. They found that the best-fit
spins show a bimodal distribution, whose peaks correspond
to the high- and low-excitation radio sources, respectively.
In particular, their model favors a trend in which the typical 
black hole spin increases slowly toward low redshift since $z\approx3$. A 
follow-up study by \cite{Martinez_Sansigre11b}, based on a similar method,
gave an even steeper evolutionary trend of
black hole spin from $z\approx1$ to $z\approx0$. In terms of
spin evolution, our results obtained here are qualitatively
consistent with those of \cite{Daly09a, Daly09b, Daly11}.  However, the stark 
differences between the results of Daly and Mart\'\i nez-Sansigre \& Rawlings 
suggest that approaches that rely on the radio 
power to deduce black hole spin should be treated with caution.

A number of semi-analytic models of galaxy formation have attempted to 
incorporate the growth of black holes and track their spin evolution.
Although the prescriptions for implementing gas fueling from large to small 
scales are uncertain and differ from study to study, there is broad consensus 
that the final black hole spin strongly depends on the assumed configuration 
of the angular momentum of the accreted gas (\citealt{Volonteri05, Berti08, 
Lagos09, Fanidakis11a, Fanidakis11b}).  The models of \cite{Lagos09} and 
\cite{Fanidakis11a} predict that the average black hole spin increases with 
mass.  At first glance, this is qualitatively consistent with our results at 
high redshifts.  However, they ascribe their results to the major mergers of 
black holes, which exclusively produce remnant holes with high spin 
(\citealt{Hughes03}).  Our calculations, by contrast, explicitly neglect black 
hole mergers and assume that mass accretion dominates black hole growth.  We
contend that this is a reasonable assumption.  Indeed, many recent studies 
cast doubt on the importance of the role played by major mergers in the 
evolution of galaxies in general (e.g., \citealt{Naab09, Taylor10, Williams11},
and references therein), and active galaxies in particular (e.g., 
\citealt{Cisternas11, Rosario11}, and references therein), especially for 
redshifts below $z\approx2$.

%
\section{Implications on Accretion Scenarios}
Two possible accretion scenarios affect black hole spins (e.g., 
\citealt{King06, Berti08, King08, Martinez_Sansigre09}).  (1) SMBHs grow 
through {\em prolonged} accretion episodes, during which the holes are quickly 
spun up to their maximum rate by capturing material with constant angular 
momentum (\citealt{Thorne74}). (2) SMBHs grow via many {\em short-lived and 
random} accretion episodes, acquiring gas with initially random orientations 
with respect to the rotation axis of the hole, which efficiently cancels out 
the net angular momentum and leads to moderately rotating holes 
(\citealt{King08, Wang09, Li10}). As in \citealt{Wang09}, we appeal to 
episodic, short-lived random accretion as the driving mechanism to explain
the intense decline of black hole spins since $z\approx1-2$.  The situation 
may be fundamentally different at higher redshifts ($z\gtrsim2$), when 
galaxies are characteristically more gas-rich and major mergers are more 
prevalent.  At these earlier epochs, it is natural to expect black hole growth 
to be more dominated by episodes of prolonged accretion, leading to high spins 
and high radiative efficiencies, particularly for high-mass systems. We 
propose that this is the fundamental explanation for the two regimes of spin 
evolution seen in our analysis (Figure \ref{fig_growth}): black holes 
spin up by prolonged accretion, possibly aided by major mergers, up to 
$z\approx2$, and thereafter spin down by episodic accretion driven by 
minor mergers and internal secular processes.

Random accretion inevitably leaves the accretion disk misaligned
with respect to the spin axis of the black hole. In this case, 
the presence of viscosity combined with Lense-Thirring 
precession will induce the inner portion of the disk to align or 
anti-align its orbital angular momentum with that of the black hole 
out to a transition radius, beyond which the disk retains its 
initial inclination (so-called Bardeen-Petterson effect; 
\citealt{Bardeen75}). The characteristic extension of the 
transition radius and the time scale for alignment or 
anti-alignment depend on the black hole mass and spin (e.g.,
\citealt{Pringle92, King05}). The Bardeen-Petterson 
effect may exert a vital influence on the spin evolution of the black hole 
(e.g., see also \citealt{Perego09}). 
We defer a complete study of the relation between spin and 
mass during multiple episodes of black hole growth to a third paper 
of this series (Y.-R. Li et al. 2012, in preparation).

%
\section{Conclusions}
We derive the mass-dependent cosmological evolution of the radiative 
efficiency for mass accretion, which, according to the standard model of 
accretion disks, can be used as a surrogate to estimate the black hole spin.
The calculated radiative efficiency generally increases with black hole mass, 
roughly as $\eta\propto M_\bullet^{0.5}$ at high redshifts ($z\gtrsim 1$), but 
the trend reverses at low redshifts, such that $\eta$ increases with 
decreasing $M_\bullet$.  High-mass black holes ($M_\bullet\gtrsim10^{8.5}\,
M_\odot$) maintain a high efficiency ($\eta\approx0.3$) at $z\approx2$, which
then declines strongly toward lower redshifts.   The evolutionary pattern for 
lower mass black holes is somewhat more complicated, but it is generally 
consistent with the radiative efficiency decreasing since $z\approx1$ in like 
manner.  Most importantly, we find that the efficiency of high-mass black 
holes begins to decline earlier than that of low-mass black holes, 
qualitatively similar to the cosmic downsizing of AGN activity.
Assuming that the radiative efficiency provides an effective indirect measure 
of the black hole spin, we propose the following picture for the spin 
evolution of SMBHs:

\begin{itemize}
 \item The evolution of the spin of the black hole tracks the growth history 
of its mass and can be characterized by two regimes: an initial phase of 
mass accumulation from prolonged accretion that spins up the hole, 
followed by a period of random, episodic accretion that spins down the hole 
toward lower redshifts.

 \item The evolution of the spin, like the global pattern of AGN activity, 
exhibits ``cosmic downsizing''. Relative to lower mass black holes, high-mass 
systems gain their masses earlier, reach the peak of their AGN activity 
earlier, and begin to spin down earlier.  Random accretion dominates their 
evolution below $z \approx 2$, whereas lower mass holes transition to this 
phase later, at $z \approx 1$.
\end{itemize}

Finally, it is worth listing the principal assumptions used in our 
calculations, on which better observational constraints would be welcomed:
(1) the scaling relation between SMBH mass and the
mass of the bulge of the host galaxy applies over the redshift range $0<z<2$,
so that the SMBH mass function can be inferred from the galaxy luminosity and 
stellar mass functions; (2) black holes gain their mass through baryon 
accretion and black hole mergers are negligible; 
(3) black hole growth mainly occurs during quasar phases characterized by
universal Eddington ratio distribution;
(4) the number of Compton-thick AGNs is comparable to that of 
Compton-thin AGNs, and they both have a similar Eddington ratio 
distribution as that of unobscured sources. Future deep multiwavelength 
surveys are awaited to test and refine these assumptions.

%
\acknowledgements{We thank the referee for helpful suggestions.
YRL acknowledges useful communications with E. S. Perego on 
the Bardeen-Petterson effect. YRL and JMW thank S. N. Zhang, H. Netzer,
C. Done, and S. Mineshige for insightful comments and suggestions, and members
of IHEP AGN group for discussions. This research is supported by NSFC-10733010, 
10821061 and 11173023, and a 973 project (2009CB824800).  The work of LCH is 
supported by the Carnegie Institution for Science.}

%
\bibliographystyle{plain}

\begin{thebibliography}{1}
\bibitem[Alexander et al.(2008)]{Alexander08}
    Alexander, D.~M., Chary, R.-R., Pope, A., et al.\ 2008, \apj, 687, 835 
\bibitem[Babi{\'c} et al.(2007)]{Babic07} 
    Babi{\'c}, A., Miller, L., Jarvis, M.~J., et al.\ 2007, \aap, 474, 755 
\bibitem[Ballantyne et al.(2006)]{Ballantyne06}
    Ballantyne, D. R., Everett, J. E. \& Murray, N. 2006, \apj, 639, 740
\bibitem[Bardeen \& Petterson(1975)]{Bardeen75} 
    Bardeen, J.~M., \& Petterson, J.~A.\ 1975, \apjl, 195, L65 
\bibitem[Berti \& Volonteri(2008)]{Berti08}
    Berti, E., \& Volonteri, M. 2008, \apj, 684, 822
\bibitem[Bongiorno et al.(2007)]{Bongiorno07}
    Bongiorno, A., Zamorani, G., Gavignaud, I., et al.\ 2007, \aap, 472, 443
 \bibitem[Cao(2010)]{Cao10} 
    Cao, X.\ 2010, \apj, 725, 388    
\bibitem[Cao \& Li(2008)]{Cao08}
    Cao, X.-W., \& Li, F. 2008, \mnras, 390, 561
\bibitem[Chokshi \& Turner(1992)]{Chokshi92}
    Chokshi, A., \& Turner, E. 1992, \mnras, 259, 421
\bibitem[Cirasuolo et al.(2010)]{Cirasuolo10} 
    Cirasuolo, M., McLure, R.~J., Dunlop, J.~S., et al.\ 2010, \mnras, 401, 1166 
\bibitem[Cisternas et al.(2011)]{Cisternas11} 
    Cisternas, M., Jahnke, K., Inskip, K.~J., et al.\ 2011, \apj, 726, 57 
\bibitem[Daddi et al.(2007)]{Daddi07}
    Daddi, E., Alexander, D.~M., Dickinson, M., et al.\ 2007, \apj, 670, 173 
\bibitem[Daly(2011)]{Daly11} 
    Daly, R.~A.\ 2011, \mnras, 395 
\bibitem[Daly(2009a)]{Daly09a} 
    Daly, R.~A.\ 2009a, \apjl, 691, L72 
\bibitem[Daly(2009b)]{Daly09b} 
    Daly, R.~A.\ 2009b, \apjl, 696, L32 
\bibitem[Davis \& Laor(2011)]{Davis11}
    Davis, S. W., \& Laor, A. 2011, \apj, 728, 98
\bibitem[Draper \& Ballantyne(2009)]{Draper09} 
    Draper, A.~R., \& Ballantyne, D.~R.\ 2009, \apj, 707, 778 
\bibitem[Dwelly \& Page(2006)]{Dwelly06} 
    Dwelly, T., \& Page, M.~J.\ 2006, \mnras, 372, 1755 
\bibitem[Elvis et al.(2002)]{Elvis02}
    Elvis, M., Risaliti, G., \& Zamorani, G. 2002, \apj, 565, L75
\bibitem[Fabian \& Iwasawa(1999)]{Fabian99} 
    Fabian, A.~C., \& Iwasawa, K.\ 1999, \mnras, 303, L34 
\bibitem[Fanidakis et al.(2011a)]{Fanidakis11a} 
    Fanidakis, N., Baugh, C.~M., Benson, A.~J., et al.\ 2011a, \mnras, 410, 53
\bibitem[Fanidakis et al.(2011b)]{Fanidakis11b} 
    Fanidakis, N., Baugh, C.~M., Benson, A.~J., et al.\ 2011b, \mnras, 419, 2979 
\bibitem[Fiore et al.(2009)]{Fiore09}
    Fiore, F., Puccetti, S., Brusa, M., et al.\ 2009, \apj, 693, 447
\bibitem[Gilli et al.(2007)]{Gilli07}
    Gilli, R., Comastri, A., \& Hasinger, G. 2007, A\&A, 463, 79
\bibitem[Gilli et al.(2010)]{Gilli10}
    Gilli, R., Comastri, A., Vignali, C., Ranalli, P., \& Iwasawa, K. 2010, in AIP Conf. Proc., X-ray
    Astronomy 2009: Present Status, Multiwavelength Approach and Future Perspectives,
    ed. A. Comastri, M. Cappi, \& L. Angelini (Melville, NY: AIP), arXiv:1004.2412
\bibitem[Greene \& Ho(2007)]{Greene07} 
    Greene, J.~E., \& Ho, L.~C.\ 2007, \apj, 667, 131 
\bibitem[Hasinger(2008)]{Hasinger08}
    Hasinger, G. 2008, A\&A, 490, 905
\bibitem[Heckman et al.(2004)]{Heckman04} 
    Heckman, T.~M., Kauffmann, G., Brinchmann, J., et al.\ 2004, \apj, 613, 109 
\bibitem[Ho(2008)]{Ho08}
    Ho, L.~C.\ 2008, \araa, 46, 475 
\bibitem[Ho(2009a)]{Ho09a} 
    Ho, L.~C.\ 2009a, \apj, 699, 626 
\bibitem[Ho(2009b)]{Ho09b} 
    Ho, L.~C.\ 2009b, \apj, 699, 638
\bibitem[Hopkins \& Hernquist(2009)]{Hopkins09} 
    Hopkins, P.~F., \& Hernquist, L.\ 2009, \apj, 698, 1550 
\bibitem[Hopkins et al.(2006)]{Hopkins06} 
    Hopkins, P.~F., Narayan, R., \& Hernquist, L.\ 2006, \apj, 643, 641 
\bibitem[Hopkins et al.(2007)]{Hopkins07}
    Hopkins, P. F., Richards, G. T., \& Hernquist, L. 2007, \apj, 654, 731
\bibitem[Hughes \& Blandford(2003)]{Hughes03}
    Hughes, S. C. \& Blandford, R. D. 2003, \apj, 585, L101
\bibitem[Kauffmann \& Heckman(2009)]{Kauffmann09} 
    Kauffmann, G., \& Heckman, T.~M.\ 2009, \mnras, 397, 135 
\bibitem[Kelly et al.(2010)]{Kelly10} 
    Kelly, B.~C., Vestergaard, M., Fan, X., et al.\ 2010, \apj, 719, 1315 
\bibitem[King et al.(2005)]{King05} 
    King, A.~R., Lubow, S.~H., Ogilvie, G.~I., \& Pringle, J.~E.\ 2005, \mnras, 363, 49 
\bibitem[King \& Pringle(2006)]{King06}
    King, A. R., \& Pringle, J. E. 2006, \mnras, 373, L90
\bibitem[King et al.(2008)]{King08}
    King, A. R., Pringle, J. E., \& Hofmann, J. A. 2008, \mnras, 385, 1621
\bibitem[Kollmeier et al.(2006)]{Kollmeier06}
    Kollmeier, J.~A., Onken, C.~A., Kochanek, C.~S., et al.\ 2006, \apj, 648, 128
\bibitem[Labita et al.(2009a)]{Labita09a} 
    Labita, M., Decarli, R., Treves, A., \& Falomo, R.\ 2009a, \mnras, 396, 1537 
\bibitem[Labita et al.(2009b)]{Labita09b} 
    Labita, M., Decarli, R., Treves, A., \& Falomo, R.\ 2009b, \mnras, 399, 2099 
\bibitem[La France et al.(2005)]{La_Franca05}
    La Franca, F., Fiore, F., Comastri, A., et al.\ 2005, \apj, 635, 864 
\bibitem[Lagos et al.(2009)]{Lagos09} 
    Lagos, C.~D.~P., Padilla, N.~D., \& Cora, S.~A.\ 2009, \mnras, 395, 625 
\bibitem[Lamastra et al.(2008)]{Lamastra08} 
    Lamastra, A., Perola, G.~C., \& Matt, G.\ 2008, \aap, 487, 109 
\bibitem[Li et al.(2011)]{Li11}
    Li, Y.-R., Ho, L. C., \& Wang, J.-M.\ 2011, \apj, 742, 33 (Paper I)
\bibitem[Li et al.(2010)]{Li10}
    Li, Y.-R., Wang, J.-M., Yuan, Y.-F., Hu, C., \& Zhang, S. 2010, \apj, 710, 878
\bibitem[Marconi et al.(2004)]{Marconi04}
    Marconi, A., Risaliti, G., Gilli, R., Hunt, L. K., Maiolino, R., \& Salvati, M. 2004, \mnras, 351, 169
\bibitem[Mart{\'{\i}}nez-Sansigre \& Rawlings(2011a)]{Martinez_Sansigre11a} 
    Mart{\'{\i}}nez-Sansigre, A., \& Rawlings, S.\ 2011a, \mnras, 414, 1937 
\bibitem[Mart\'\i nez-Sansigre \& Rawlings(2011b)]{Martinez_Sansigre11b} 
    Mart\'\i nez-Sansigre, A., \& Rawlings, S.\ 2011b, \mnras, 418, L84  
\bibitem[Mart\'{i}nez-Sansigre \& Taylor(2009)]{Martinez_Sansigre09}
    Mart\'{i}nez-Sansigre, A., \& Taylor, A. M. 2009, 692, 964
\bibitem[Martini(2004)]{Martini04}
    Martini, P. 2004, in Carnegie Observatories Astrophysics Series, Vol. 1: Coevolution of Black Holes and Galaxies, ed. L. C. Ho (Cambridge: Cambridge Univ. Press), 169
\bibitem[Merloni \& Heinz(2008)]{Merloni08} 
    Merloni, A., \& Heinz, S.\ 2008, \mnras, 388, 1011 
\bibitem[Naab et al.(2009)]{Naab09} 
    Naab, T., Johansson, P.~H., \& Ostriker, J.~P.\ 2009, \apjl, 699, L178 
\bibitem[Netzer et al.(2007)]{Netzer_etal07} 
    Netzer, H., Lira, P., Trakhtenbrot, B., Shemmer, O., \& Cury, I.\ 2007, \apj, 671, 1256 
\bibitem[Netzer \& Trakhtenbrot(2007)]{Netzer07} 
    Netzer, H., \& Trakhtenbrot, B.\ 2007, \apj, 654, 754 
\bibitem[Peacock(1999)]{Peacock99}
    Peacock, J. A. 1999, Cosmological Physics (Cambridge: Cambridge Univ. Press)
\bibitem[Perego et al.(2009)]{Perego09} 
    Perego, A., Dotti, M., Colpi, M., \& Volonteri, M.\ 2009, \mnras, 399, 2249 
\bibitem[Pringle(1992)]{Pringle92} 
    Pringle, J.~E.\ 1992, \mnras, 258, 811 
\bibitem[Raimundo et al.(2011)]{Raimundo11} 
    Raimundo, S.~I., Fabian, A.~C., Vasudevan, R.~V., Gandhi, P., \& Wu, J.\ 2011, 
    \mnras, 419, 2529 
\bibitem[Rosario et al.(2011)]{Rosario11} 
    Rosario, D.~J., Mozena, M., Wuyts, S., et al.\ 2011, \apj, in press (arXiv:1110.3816) 
\bibitem[Salpeter(1964)]{Salpeter64} 
    Salpeter, E.~E.\ 1964, \apj, 140, 796 
\bibitem[Schulze \& Wisotzki(2010)]{Schulze10} 
    Schulze, A., \& Wisotzki, L.\ 2010, \aap, 516, A87
\bibitem[Shankar(2009)]{Shankar09}
    Shankar, F. 2009, New Astron. Rev., 53, 57
\bibitem[Shankar et al.(2010)]{Shankar10}
    Shankar, F., Crocce, M., Miralda-Escud\'{e}, J., Fosalba, P., \& Weinberg, D. H. 2010, \apj, 718, 231
\bibitem[Shankar et al.(2004)]{Shankar04}
    Shankar, F., Salucci, P., Granato, G. L., De Zotti, G., \& Danese, L. 2004, \mnras, 354, 1020
\bibitem[Shankar et al.(2009)]{Shankar_etal09}
    Shankar, F., Weinberg, D. H., \& Miralda-Escud\'{e}, J. 2009, \apj, 690, 20
\bibitem[Shapiro(2005)]{Shapiro05}
    Shapiro, S. L. 2005, \apj, 620, 59
\bibitem[Shen et al.(2008)]{Shen08}
    Shen, Y., Greene, J. E., Strauss, M. A., Richards, G. T., \& Schneider, D. P. 2008, \apj, 680, 169
\bibitem[Sikora et al.(2007)]{Sikora07}
    Sikora, M., Stawarz, L., \& Lasota, J.-P. 2007, \apj, 658, 815
\bibitem[Small \& Blandford(1992)]{Small92}
    Small, T. A., \& Blandford, R. D. 1992, \mnras, 259, 725
\bibitem[So{\l}tan(1982)]{Soltan82}
    So{\l}tan, A. 1982, \mnras, 200, 115
\bibitem[Taylor et al.(2010)]{Taylor10} 
    Taylor, E.~N., Franx, M., Glazebrook, K., et al.\ 2010, \apj, 720, 723 
\bibitem[Thorne(1974)]{Thorne74}
    Thorne, K. S. 1974, \apj, 191, 507
\bibitem[Treister et al.(2010)]{Treister10}
    Treister, E., Natarajan, P., Sanders, D. B., Urry, C. M., Schawinski, K., \& Kartaltepe, J. 2010, Science, 328, 600
\bibitem[Treister \& Urry(2006)]{Treister06}
    Treister, E., \& Urry, C. M. 2006, \apj, 652, L79
\bibitem[Treister et al.(2009)]{Treister09}
    Treister, E., Cardamone, C.~N., Schawinski, K., et al.\ 2009, \apj, 706, 535
\bibitem[Ueda et al.(2003)]{Ueda03}
    Ueda, Y., Akiyama, M., Ohta, K., \& Miyaji, T. 2003, \apj, 589, 886
\bibitem[Wang et al.(2008)]{Wang08} 
    Wang, J.-M., Chen, Y.-M., Yan, C.-S., \& Hu, C.\ 2008, \apjl, 673, L9 
\bibitem[Wang et al.(2006)]{Wang06}
    Wang, J.-M., Chen, Y.-M., \& Zhang, F. 2006, \apj, 647, L17
\bibitem[Wang et al.(2009)]{Wang09}
    Wang, J.-M., Hu, C., Li, Y.-R., et al.\ 2009, \apjl, 697, L141 
\bibitem[Williams et al.(2011)]{Williams11} 
    Williams, R.~J., Quadri, R.~F., \& Franx, M.\ 2011, \apjl, 738, L25 
\bibitem[Vestergaard \& Osmer(2009)]{Vestergaard09} 
    Vestergaard, M., \& Osmer, P.~S.\ 2009, \apj, 699, 800 
\bibitem[Volonteri et al.(2008)]{Volonteri08} 
    Volonteri, M., Lodato, G., \& Natarajan, P.\ 2008, \mnras, 383, 1079 
\bibitem[Volonteri et al.(2005)]{Volonteri05}
    Volonteri, M., Madau, P., Quataert, E., \& Rees, M. J. 2005, \apj, 667, 704
\bibitem[Volonteri et al.(2007)]{Volonteri07}
    Volonteri, M., Sikora, M., \& Lasota, J.-P. 2007, \apj, 667, 704
\bibitem[Xu \& Cao(2010)]{Xu10} 
    Xu, Y.-D., \& Cao, X.\ 2010, \apj, 716, 1423 
\bibitem[Yu et al.(2005)]{Yu05} 
    Yu, Q., Lu, Y., \& Kauffmann, G.\ 2005, \apj, 634, 901 
\bibitem[Yu \& Tremaine(2002)]{Yu02}
    Yu, Q., \& Tremaine, S. 2002, \mnras, 335, 965
\end{thebibliography}

\end{document}